\documentclass[a4paper]{article}

\usepackage{INTERSPEECH2021}
\usepackage{booktabs}
\usepackage{makecell}
\usepackage{graphicx}
\usepackage{flushend}  % make reference columns balanced
\usepackage{xcolor}

\title{A Hands-on Comparison of DNNs for Dialog Separation \\ Using Transfer Learning from Music Source Separation}
\name{Martin Strauss$^1$, Jouni Paulus$^{1,2}$, Matteo Torcoli$^2$, Bernd Edler$^1$}
\address{
  $^1$International Audio Laboratories Erlangen\textsuperscript{*}\thanks{\textsuperscript{*}A joint institution of the Friedrich-Alexander-University Erlangen-N\"{u}rnberg (FAU) and Fraunhofer IIS, Germany.}, Germany\\
  $^2$Fraunhofer Institute for Integrated Circuits IIS, Erlangen, Germany}
\email{\{martin.strauss, bernd.edler\}@audiolabs-erlangen.de, \\ \{jouni.paulus, matteo.torcoli\}@iis.fraunhofer.de}

\begin{document}

\maketitle
\begin{abstract}  % max 200 words
This paper describes a hands-on comparison on using state-of-the-art music source separation deep neural networks (DNNs) before and after task-specific fine-tuning for separating speech content from non-speech content in broadcast audio (i.e., dialog separation).
The music separation models are selected as they share the number of channels (2) and sampling rate (44.1\,kHz or higher) with the considered broadcast content, and vocals separation in music is considered as a parallel for dialog separation in the target application domain.
These similarities are assumed to enable transfer learning between the tasks.
Three models pre-trained on music (Open-Unmix, Spleeter, and Conv-TasNet) are considered in the experiments, and fine-tuned with real broadcast data.
The performance of the models is evaluated before and after fine-tuning with computational evaluation metrics (SI-SIRi, SI-SDRi, 2f-model), as well as with a listening test simulating an application where the non-speech signal is partially attenuated, e.g., for better speech intelligibility.
The evaluations include two reference systems specifically developed for dialog separation.
The results indicate that pre-trained music source separation models can be used for dialog separation to some degree, and that they benefit from the fine-tuning, reaching a performance close to task-specific solutions.
\end{abstract}
\noindent\textbf{Index Terms}: dialog separation, deep neural network, transfer learning, fine-tuning

\section{Introduction}
A common complaint about the audio mix in TV-shows or movies is the low intelligibility of the spoken dialogs as a consequence of loud background music or sound effects~\cite{armstrong:2016}.
The right balance between dialog and background depends on many personal factors, such as hearing acuity, listening environment, and individual taste~\cite{geary2020loudness}.
The ideal solution is when a person can adjust the volume of the dialog content individually from the background sounds according to their personal preferences and needs.
Such end-user interactivity is enabled by object-based audio systems, such as MPEG-H Audio~\cite{simon2019mpeg}, where the dialog and background can be transported as distinct objects and mixed at the receiver.
The required audio tracks can be saved in new productions, but there exists a significant amount of archived content for which only the final audio mixture is available.
For obtaining the object of these mixtures, source separation methods are employed~\cite{paulus2019source, Lim19-BMSB,uhle2008speech}.
Although audio source separation is a major research topic since many years~\cite{vincent_book}, techniques applicable for broadcast content are not easily available. Among others, the main reasons for this can be identified as:
1) Broadcast content is usually multi-channel, e.g., stereo, which makes the separation process different from the single-channel methods in the literature.
2) Broadcast content needs to be of high perceptual quality and requires high sampling rates (usually 48\,kHz).
3) Real broadcast content with clean reference signals that can be used for training is generally not available to researchers.

Approaches that specifically aim at this application are still rare.
Examples of solutions with non-DNN (deep neural network) signal processing include \cite{geiger_2015}~extracting a phantom center channel containing the dialog from a stereo mix, \cite{master2020dialog}~where source separation based on spatio-level filtering is combined with a classifier induced gating, and an earlier work from the authors~\cite{paulus2019source} using a late-fusion of multiple parallel methods.
Although the proposed systems are of relatively low computational complexity and show favourable subjective performance, the development of their overall performance has somewhat stagnated.

Approaches using DNNs are successful in many tasks, and can be considered as the de-facto state-of-the-art also in source separation of speech signals~\cite{Wang_survey}.
Even though significant progress has been made in both, single-channel~\cite{conv_tasnet, Ephrat_audio_visual} and multi-channel speech enhancement~\cite{Li19-WASPAA, Sekiguchi_transaction, Chang_transactions, Luo19-MSC, Pandey18-ICASSP}, the proposed systems are still largely impractical for an application in the broadcast environment mainly being trained on low sampling rates or supporting only single input channel.
The method of~\cite{Mack20-SPL} uses an LSTM for predicting complex-valued 2D filters for speech separation with the benefit over instantaneous filters that these can handle also missing input data to some degree.
Many of the speech separation models developed at 8 or 16\,kHz sampling rates suffer from a model size explosion when applied on the sampling rate required by the application, because of cubical or worse dependency on the frequency resolution. The authors in \cite{Li19-WASPAA}~addresses this problem by sharing the long-short-term-memory (LSTM) parameters across the frequencies.

Making an efficient use of the limited training data is important and in~\cite{Stoller18-ICASSP} the authors describe a Generative Adversarial Network based system for this, but demonstrate it only for music separation. Still, this might be applicable also for dialog separation in broadcast content.

At the same time, progress in music source separation has been made by developing models trained on stereo data with high sampling rates, and some of these models are able to separate the vocals from the rest of the musical piece~\cite{Cano2019}.
Assuming that speech and singing are close enough as signals and transferring knowledge from those pre-trained music networks towards dialog separation, could help to overcome multiple of the aforementioned issues.

This paper investigates the application of music source separation models and applying transfer learning for solving the dialog separation task.
We fine-tune three state-of-the-art neural networks (NN): Open-Unmix~\cite{open_unmix}, Spleeter~\cite{spleeter2020}, and Conv-TasNet~\cite{defossez2019music}, with real-world broadcast material and compare their performance against two systems developed specifically for the task.
The performance is evaluated using both computational measures (Scale-Invariant BSSEval, 2f-model), as well as a listening test simulating the intended application of enhancing the relative level of the dialog content in the mixture.
% is this result statement needed in intro?
%\textcolor{red}{The experimental outcome demonstrates that the models trained for music tasks may be applied also on other tasks to some degree, and that they benefit from refinement training with task-specific data leading some of the adapted models to perform competitively compared to solutions specifically aimed at this application.}

\section{Problem Formulation}
Let us consider a stereo audio mixture $\textbf{y}\in \mathbb{R}^{N \times 2}$ of length $N$, composed of a summation of the two components $\textbf{x}\in \mathbb{R}^{N \times 2}$ and $\textbf{b}\in \mathbb{R}^{N \times 2}$, i.e.,
\begin{equation}
    \textbf{y}=\textbf{x}+\textbf{b}.
\end{equation}
Signal \textbf{x} represents the dialog component consisting of one or more unknown talkers. The background signal \textbf{b} potentially includes a wide range of sounds, e.g., music and sound effects, but also ambient nature or crowd noise. The aim of dialog separation is to extract the single components $\hat{\textbf{x}}$ and $\hat{\textbf{b}}$, with them being close to the underlying $\textbf{x}$ and $\textbf{b}$, so that they can be remixed with an adapted relative level of the dialog, e.g., by a background attenuation factor $\mu$:
\begin{equation}
    \hat{\textbf{y}}= \hat{\textbf{x}}+\mu \hat{\textbf{b}}.
\end{equation}
As a result, the audibility and potentially also intelligibility of the dialog in $\hat{\textbf{y}}$ is increased, ideally without decreasing the overall perceptual quality of the audio signal.

\section{Separation Methods}
In the following we consider three DNNs applicable for separating vocals from music with minor modification to match our data and fine-tuning them with dialog-specific data (Sec.\,\ref{sec:UNM} -- \ref{sec:ConvT}).
The performance of these is compared with two solutions specifically intended for dialog separation (Sec.\,\ref{sec:DS}).

\begin{table*}[ht]
\renewcommand{\arraystretch}{1.2}
\centering
\caption{Experimental results using the computational evaluation metrics. The \textbf{Input} column describes the values of the original mix. The columns containing \textbf{base} and \textbf{transfer} represent the corresponding methods without and with fine-tuning using broadcast content. \textbf{Conv-TasNet retrained} shows the results of training Conv-TasNet completely from scratch (i.e. random weights initialization instead of pre-trained weights). In the first two rows, the mean of the absolute SI-SDR and SI-SIR values are displayed. The SI-SDRi and SI-SIRi are computed as the average of the per-item improvement. The last row shows the 2f-model score of the analysed methods.}
\resizebox{\textwidth}{!}{%
\begin{tabular}{lccccccccccc}%\toprule
\textbf{Method} & \textbf{Input} & \multicolumn{1}{l}{\textbf{\makecell{IRM}}}& \multicolumn{1}{l}{\textbf{\makecell{Conv-TasNet\\ retrained}}} &\multicolumn{1}{l}{\textbf{\makecell{UMX \\ base}}} & \multicolumn{1}{l}{\textbf{\makecell{Spleeter \\ base}}} & \multicolumn{1}{l}{\textbf{\makecell{Conv-TasNet\\base}}} & \multicolumn{1}{l}{\textbf{\makecell{UMX \\ transfer}}} & \multicolumn{1}{l}{\textbf{\makecell{Spleeter \\transfer}}} & \multicolumn{1}{l}{\textbf{\makecell{Conv-TasNet\\ transfer}}} &\multicolumn{1}{l}{\textbf{\makecell{IDC}}} & \multicolumn{1}{l}{\textbf{\makecell{DNN-DS}}} \\ \midrule
SI-SDR\hphantom{-} [dB] & 6.5 $\pm$ 7.0 & 16.8 $\pm$ 5.6 & 15.5 $\pm$ 5.8 & 10.9 $\pm$ 5.7 & 12.0 $\pm$ 5.9 & 13.8 $\pm$ 5.2 & 13.0 $\pm$ 5.7 & 14.4 $\pm$ 6.1 & \bf{16.8 $\pm$ 5.1} & \hphantom{7}5.1 $\pm$ 5.5 & 14.5 $\pm$ 5.9\\
SI-SIR \hphantom{-i}[dB] & 6.5 $\pm$ 7.1 & 22.4 $\pm$ 6.6 & 23.9 $\pm$ 6.1 & 18.8 $\pm$ 7.4 & 18.4 $\pm$ 6.1 & 19.1 $\pm$ 6.4 & 20.4 $\pm$ 6.8 & 23.5 $\pm$ 7.0 & \bf{28.4 $\pm$ 6.8} & 20.6 $\pm$ 8.7 & 23.6 $\pm$ 7.5 \\\midrule
SI-SDRi [dB] &  & 10.3 $\pm$ 4.2 & \hphantom{7}9.0 $\pm$ 3.9 & \hphantom{7}4.4 $\pm$ 5.1 & \hphantom{7}5.5 $\pm$ 3.7 & \hphantom{7}7.3 $\pm$ 4.2 & \hphantom{7}6.5 $\pm$ 4.6 & \hphantom{7}7.9 $\pm$ 4.1 & \bf{10.3 $\pm$ 4.2} & -1.5 $\pm$ 4.6 & \hphantom{7}8.0 $\pm$ 4.1 \\
SI-SIRi \hphantom{-}[dB]& & 15.9 $\pm$ 6.9 &  17.4 $\pm$ 7.5 & 12.3 $\pm$ 6.9 & 12.0 $\pm$ 6.6 & 12.6 $\pm$ 7.7 & 13.9 $\pm$ 7.5 & 17.1 $\pm$ 8.3 & \bf{21.9 $\pm$ 8.8} & 14.1 $\pm$ 6.6 & 17.1 $\pm$ 8.3 \\\midrule
%SI-SAR [dB] & 12.24 & 15.42 & 17.32 & 14.56 & 15.42 & 17.32 & 5.37 & 15.61 \\ \midrule
2f-model & 20.6 $\pm$ 11.9 & \bf{70.1 $\pm$ 9.3} &  34.3 $\pm$ 6.9 & 31.0 $\pm$ 8.7 & 33.9 $\pm$ 11.6 & 36.8 $\pm$ 11.8 & 35.1 $\pm$ 9.7 & 38.6 $\pm$ 13.4 & 46.4 $\pm$ 12.2 & 23.8 $\pm$ 8.2 & 35.0 $\pm$ 12.4 \\ %\bottomrule
\end{tabular}}
\label{tab:results}
\end{table*}

\subsection{Open-Unmix (UMX)}
\label{sec:UNM}
UMX~\cite{open_unmix} was created as the reference system in the SiSEC 2018 evaluation campaign~\cite{SiSEC18}. The model operates in Short-time Fourier transform (STFT) domain and is a DNN-based on a three-layer bidirectional LSTM with fully-connected encoding and decoding layers to estimate the separation masks.
One network is trained for each target source (vocals, bass, drums, other) using the mean-squared-error loss between the target and the separated magnitude spectrograms.
During inference, a multi-channel Wiener filter post-processing step is applied. For this work the model trained on the uncompressed version of the MUSDB18 (MUSDB18-HQ)~\cite{MUSDB18} dataset was used.
Note, that although a specific speech enhancement version of this model is available, it was not further considered in this work, since it was trained on single-channel audio and at 16\,kHz sampling rate.
We selected the model with vocals target for our experiments.
The model has approximately 8.9\,M trainable parameters.

\subsection{Spleeter}
Spleeter~\cite{spleeter2020} is a U-Net like convolutional NN operating in STFT-domain with 12 layers.%: 6 layers for encoding and 6 for decoding.
Although each target source has its own network, these are optimized jointly by considering the combined loss.
The training target is a soft mask for each source based on the corresponding magnitude spectrograms and the $L_{1}$-loss.
 Also here, during inference an additional Wiener filtering is performed.
The network was trained on a large internal dataset of Deezer.
According to the authors, the network was only trained on a bandwidth up to 11\,kHz for performance reasons.
Still, we found that separation up to 22\,kHz works reasonably well.
For the experiments, we created our own PyTorch implementation of the model using the pre-trained weights of the 2-source-model and fine-tuning it on our data.
Additionally, the batch normalization layers were frozen for the fine-tuning, because this improved the stability.
%we noticed that without this step the network spends more time on adapting to the new statistics of the training data.
The model has approximately 19.6\,M trainable parameters.

\subsection{Conv-TasNet for music source separation}
\label{sec:ConvT}
The Conv-TasNet model for music source separation~\cite{defossez2019music} is a multi-channel version of the Conv-TasNet for speaker-speaker separation~\cite{conv_tasnet}.
%In the music version the model parameters were adapted to match the original receptive field of 1.5\,s, but for 44.1\,kHz.
The network operates in time domain by learning soft masks for each source, based on a learnt linear encoder-decoder representation using the $L_{1}$-loss of the time-domain signals.
The model is fully-convolutional using dilated 1-D convolutional blocks to incorporate temporal information.
%During inference the network creates 8\,s long chunks which are processed independently and concatenated.
There was no openly available model trained for two target components.
For this reason we employed the model for four sources trained on 150 additional songs.
The separation result of the three non-vocal sources was summed together to create a background component before being forwarded to the loss computation.
Due to GPU memory constraints, the batch size had to be reduced to 1 for the fine-tuning.
We are aware that the paper~\cite{defossez2019music} proposes Demucs as an alternative for source separation with better results.
However, we were unable to fit it to our GPU, which is why it was not considered further.
Note, that for the remainder of this paper we use "Conv-TasNet" to refer to the music separation model.
The model has approximately 11\,M trainable parameters.

\subsection{Systems specifically developed for dialog separation}
\label{sec:DS}
\subsubsection{Audionamix Instant Dialogue Cleaner (IDC)}
IDC~\cite{IDC} is a commercially available plugin for dialog separation, e.g., for production or broadcasting engineers.
According to the Audionamix website, it is built on DNNs and operates in real-time.
For our study, the estimated dialog component was obtained by setting the plugin to maximum background attenuation (i.e., background gain was set to $-\infty$\,dB, speech gain was set to 0\,dB, and the \textit{Strength} parameter was set to 0).

\subsubsection{DNN Dialog Separation (DNN-DS)}
The second comparison method DNN-DS is the signal separation core of a solution under development.
The full system has been employed recently in field tests in Germany~\cite{dnn_ds}.
Since the pre- and post-processing steps of the full system have a significant effect on the output quality and they have been tuned specifically for this method,  they were disabled for allowing a better comparison of plain network performance.
The method uses a fully-convolutional structure in STFT-domain with a receptive field of approximately 1\,s.
The model has approximately 370'000 trainable parameters, and for this study, it was trained using the same data as is used for the fine-tuning of the candidate systems.

\subsection{Data}
The data used in this work is comprised of real-world broadcast material, hand-cleaned to obtain clean reference components by retaining signal parts with only clean speech (or silence) or non speech-like sounds in the respective temporally aligned stems.

All items were in stereo at 48\,kHz sampling rate.
The content ranged from nature documentaries to sports and drama.
The data contained both female and male talkers in multiple languages, though the majority was in German.
As a pre-processing step the data was downsampled to 44.1\,kHz to match the network requirements.
Then the training items in the dataset were cut into 15\,s chunks. The dataset was separated into a training (15\,h), a validation (1.45\,h) and a test set (0.9\,h).
%It was ensured, that the talkers in the dialog components were not mixed in between the sets.

\subsection{Training strategy}
The input training samples were randomly selected from each clip.
Since the subject networks utilize a different amount of input data (UMX 6\,s, Conv-TasNet 2.8\,s and Spleeter 12\,s), we wanted to make sure that each model was trained on a similar amount of data.
Therefore, per epoch the items of UMX and Conv-TasNet were sampled two and four times respectively to match the duration of input data for Spleeter.

A learning rate decay by a factor of 0.3 and a patience of 5 epochs was included into training.
The maximum number of epochs was limited to 100.
To prevent overfitting, an additional early-stopping mechanism was implemented, stopping the training if the validation loss did not decrease for 10 consecutive epochs.
All data augmentation was turned off, since the methods used varied in between the different networks.

\section{Evaluation Methods}
The performance of the models is evaluated before and after fine-tuning with computational evaluation metrics (Sec.\,\ref{sec:obj}), as well as with subjective evaluation via a listening test (Sec.\,\ref{sec:subj}).

\subsection{Computational evaluation metrics}
\label{sec:obj}
To quantify the outcome of our experiments we compute the improvement in Scale-Invariant Signal-to-Distortion-Ratio (SI-SDRi) and Signal-to-Inference-Ratio (SI-SIRi)~\cite{si_sdr} in dB between the separated and reference components of the test set.
%These are common metrics to rate the raw performance of separation models.
In addition, the average SI-SDR and SI-SIR values of the input mix as well as the corresponding output values of the separated dialog components are reported.
As a further metric we computed the 2f-model score~\cite{2f}.
It is a value in the range 0 -- 100 approximating the result of a MUSHRA listening test, and it has been shown to correlate well with perceptual audio quality for source separation~\cite{Torcoli21-TASLP}.
Since oracle ideal ratio mask (IRM) is often considered as the training target for source separation DNNs, we computed the metrics for it as an additional orientation point.
The IRM results were obtained using a sine window of 43\,ms, 50\% overlap and an exponent of 2 defining a Wiener filter-like separation mask.
%The selected music separation models are evaluated before (base) and after (transfer) the refinement.

\begin{figure*}[htb]
\centering
\begin{minipage}[b]{0.88\textwidth}
\includegraphics[width=1\textwidth]{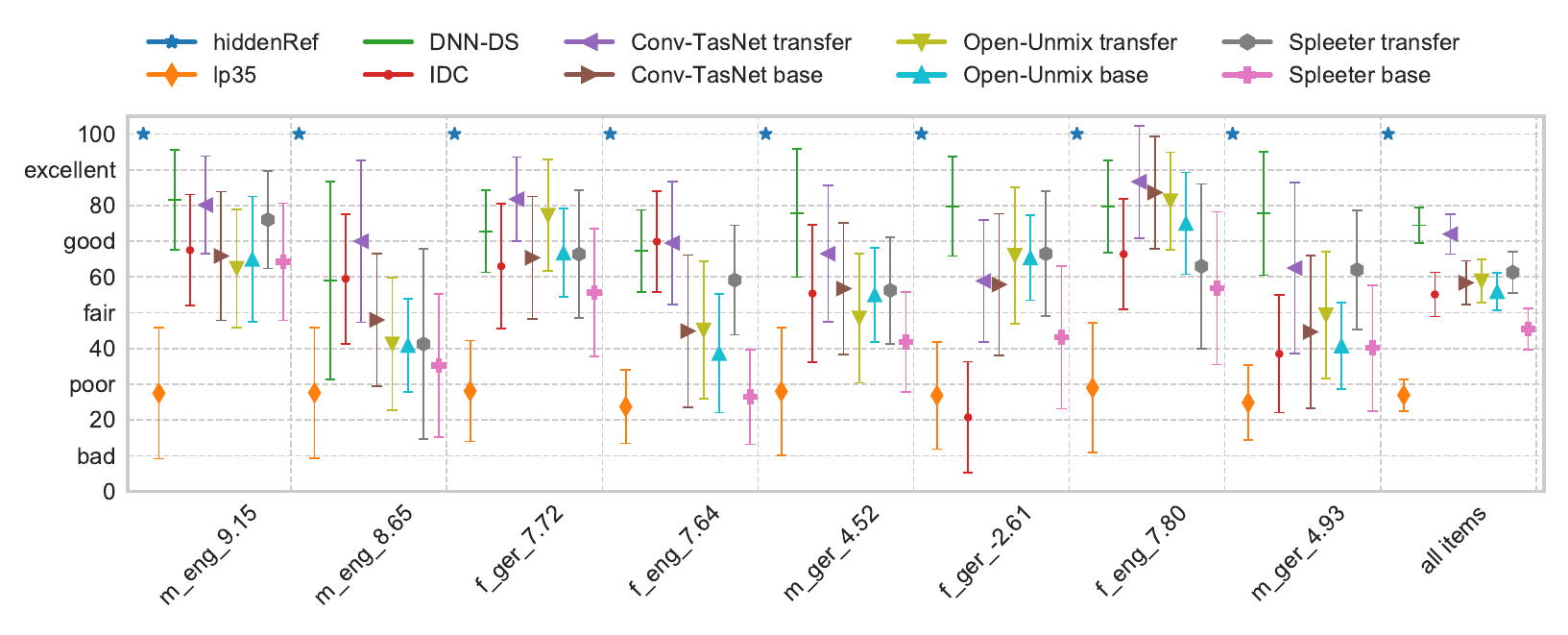}
\end{minipage}
\caption{Listening test results. Mean and 95\% confidence interval (using Student's t-distribution, 8 listeners) are depicted per item and test condition. The test items include female (\textbf{f}) and male (\textbf{m}), as well as German (\textbf{ger}) and English (\textbf{eng}) samples, and different input SI-SIR (reported at the end of the each sample name). On the right, the average results are displayed.
%Next to the tested systems \textbf{hiddenRef} described the reference condition and \textbf{lp35} the low anchor condition obtained by a 3.5\,kHz low-pass filter.
(Best viewed in colors)}
%\vspace{-0.6cm}
\label{fig:mushra}
\end{figure*}

\subsection{Listening test}
\label{sec:subj}
A listening test was conducted with a subset of 8 items to evaluate the perceptual quality.
The items were selected to obtain 4 items with female and 4 with male talkers each, as well as 4 in German and 4 in English language; 4 items featured music as background, while 4 items featured noisy background, e.g., atmospheres from forest, seaside, or football stadium.
Each item was 8\,s long.
The computational evaluation was repeated on the selected items, confirming that the selection was not biased towards a specific model.
The separated components were mixed to simulate a dialog enhancement application scenario by:
\begin{enumerate}
    \item Create a reference condition by attenuating the background component by 12\,dB compared to the input mix.
    \item Calculate the integrated loudness (as per~\cite{loudness_standart}) of the reference background considering only the non-dialog parts, with gating deactivated.
	\item For each separation method tested (i.e., test condition), obtain the background component by subtracting the estimated dialog component from the input mixture.
    \item For each test condition, determine a background attenuation to obtain the same background loudness during dialog inactivity as in the reference condition. Maximal attenuation is 120\,dB.
    \item Normalize each item to -23\,LUFS integrated loudness (gating deactivated).
\end{enumerate}

The test was conducted following the MUSHRA methodology~\cite{mushra2015}.
Including a 3.5\,kHz low-pass anchor and the hidden reference, the test consisted of 10 conditions.
Note, that the IRM was not included as a condition into the listening test, since it is an ideal method, not available in practice.
Due to the COVID-19 restrictions, the test had to be conducted at home office, each participant with their own computer and headphones.
The test participants were fellow post-graduate students and colleagues of the authors, with various levels of experience in subjective evaluation of audio and at least a basic knowledge of German and English.
The authors did not participate in the test.
The test results from 14 listeners were collected, and 6 participants were post-screened because they did not always rate the hidden reference as the best condition.

\section{Experimental results}
\subsection{Computational evaluation metrics}
The results of the computational evaluation metrics are displayed in Table\,\ref{tab:results}.
All the fine-tuned models outperform the respective base versions showing a benefit introduced by the task-specific fine-tuning step and suggesting that transfer learning from music separation towards dialog separation is possible.
%%%%%%%%%%%%%%%%%%%
% Does it?
%%%%%%%%%%%%%%%%%%%

UMX has the lowest SI-SDRi and 2f-model scores both before and after fine-tuning.
One possible explanation for this is the smaller amount of training data UMX uses in the original training.
Considering SI-SIRi, UMX shows a better performance for the base model than the base version of Spleeter but, after fine-tuning the Spleeter results exceed UMX by a large margin.
Conv-TasNet consistently shows better results than the other two methods before and after fine-tuning.
The margin even increases after fine-tuning suggesting that Conv-TasNet benefits from this more than the other models.
IDC shows a better performance than the base networks and the fine-tuned UMX considering SI-SIRi, but after fine-tuning it is being outperformed by Spleeter and Conv-TasNet.
The SI-SDRi for IDC is negative indicating that the processing induces significant artifacts.
Inspecting some of the output signals confirms this and reveals that occasionally the dialog components are missed entirely.
It has to be noted that IDC operates causally in real-time, and therefore has more challenging processing constraints than the other methods.
The performance of DNN-DS is significantly better than both versions of UMX, close to the fine-tuned Spleeter, and is outperformed by Conv-TasNet in the computational metrics.
In average SI-SIRi, Conv-TasNet exceeds DNN-DS by 4\,dB indicating a better dialog/background separation.
IRM is outperformed by Spleeter, Conv-TasNet and DNN-DS in SI-SIRi while showing the best 2f-model results by a large margin.
%Similar ordering of methods can be observed in the 2f-model score.
The beneficial effect of transfer learning can be further highlighted by considering a Conv-TasNet model that was trained from scratch (random weights initialization) using the same parameters and data. The metrics in Table\,\ref{tab:results} indicate that training from scratch can outperform the respective base model for the task at hand, but performs worse than the fine-tuned version pretrained on music. Similar behaviour was observed with Spleeter and UMX.

\subsection{Listening test}
The results from the listening test are displayed in Figure\,\ref{fig:mushra}.
It can be seen that DNN-DS reaches the best performance with a slight advantage over Conv-TasNet.
The advantage is particularly noticeable in the test items with lower input SI-SIR ($\mathrm{<}$5\,dB), while the networks are on par or Conv-TasNet takes the lead for higher input SI-SIR ($\mathrm{\geq}$5\,dB).
The overall slight advantage of DNN-DS is contrary to the computational evaluation where Conv-TasNet was ahead in all metrics, reminding that even state-of-the-art computational evaluation measures may give results different from subjective evaluation~\cite{Torcoli21-TASLP}.
One thing to note is that Conv-TasNet uses almost 30 times more trainable parameters than DNN-DS to reach this performance.
Furthermore, we assume that Conv-TasNet benefits significantly from the remixing in the application. Inspecting some separated dialog samples reveals quite audible clipping-like artifact.
Remixing the dialog component with an attenuated version of the background component appears to hide these artifacts reasonably well.
A point for future work is to test the method proposed in~\cite{Bahmaninezhad19-INTERSPEECH} replacing the learned encoder and decoder with STFT.
This should both reduce the model size considerably and address the distinctive timbral artifact.\\
\indent IDC performs better in the listening test than the base UMX and the base Spleeter, but stays behind the other methods.
This confirms the low scores observed in the computational evaluation metrics.
The fine-tuned Spleeter network shows still the second best results of the music models, but with an increased margin from the best performing models.
This could be explained by phantom speech or chorusing-like artifacts that Spleeter introduces.
This effect is even emphasized in the base Spleeter version, showing the lowest listening test scores of all music networks. The categorization of UMX is largely confirmed by the listening test.

\section{Conclusion}
This paper described a comparison of applying three state-of-the-art music separation DNN models for dialog separation before and after task-specific fine-tuning.
When evaluated in the remixing application, the considered DNNs range between perceptual quality levels described as fair and good.
This shows that dialog separation can be solved to a degree using transfer learning from music separation, with competitive performance compared to solutions specifically developed for the task.
The results also showed that the methods all benefit to varying degrees from fine-tuning training with task-specific material.

%\section{Acknowledgements}
%The authors would like to thank all participants of the listening test for their time.

\bibliographystyle{IEEEtran}

\bibliography{mybib}

% use "bibtex -min-crossrefs=99" to avoid creating new entries for crossref'd conferences

\end{document}